\documentclass{ws-p8-50x6-00}

\begin{document}

\title{HIGH $P_T$ PHYSICS WITH THE STAR EXPERIMENT AT 
  RHIC\footnote{Talk given at APS Centennial Meeting, Atlanta, GA, 
   March 1999}}

\author{K. TURNER}

\address{Physics Department, Brookhaven National Lab. \\
         Upton, NY 11973, USA}


\author{(for the STAR Collaboration)}


\maketitle


\abstracts{
The STAR experiment at RHIC is a TPC-based, general purpose 
 detector designed to obtain charged particle spectra, with an 
 emphasis on hadrons over a large phase space.
An electromagnetic calorimeter provides measurement of $e$'s, $\gamma$'s,
  $\pi^0$'s and jets.
Data-taking with $Au + Au$ collisions at $\sqrt s = 200 ~\rm{GeV/c^2}$
  begins in Fall 1999.
The STAR experiment's investigation of techniques and signals using
  hard probes to study the high energy-density matter at RHIC and to
  search for quark-gluon plasma formation will be described.
}


\section{Introduction}

 The Relativistic Heavy Ion Collider
 (RHIC) 
 at BNL will provide collisions of ions from $p$ to $Au$ at 
 $\sqrt s $ up to $500 ~{\rm GeV/c^2}$ ($p$ beams) and
       $200 ~{\rm GeV/c^2}$ ($Au$ beams) beginning in Fall 1999.
The STAR experiment\cite{star_cdr} is designed to study the
  high energy-density nuclear matter produced in these collisions
  and to search for the phase transition
  to a quark-gluon plasma (QGP).
The QGP is a deconfined state of quarks and gluons, predicted by 
  quantum chromodynamics (QCD) to exist  at high energy-densities.
One facet of the STAR experimental  program is to use
  high transverse momentum ($p_T$) production to probe the dense
  matter produced at RHIC.
In this paper, the motivation for studying high-$p_T$
  production in heavy ion collisions, predicted signatures of QGP 
  using hard probes and the planned STAR measurements are described.


Measurements of  high-$p_T$ production from high energy collisions
  allow small distances,
  and therefore the earliest times after the collision, to be probed.
The high-$p_T$ partons 
  retain information about the collision during hadronization.
High-$p_T$ production from hadron-hadron collisions 
 has been shown to be well-described by perturbative QCD 
 (pQCD)\cite{qcd_works}.

At RHIC,  
  it has been  estimated that up to 50\% of the transverse energy
  produced is due to partonic processes\cite{xnwang_92}.
Therefore, pQCD predictions become viable for the first time
  in relativistic heavy ion collisions.
Incorporating standard nuclear effects such as  nuclear modification of
  the parton distribution functions\cite{emc_pdf} 
  and the Cronin effect\cite{cronin} into the pQCD
  calculations lead to  accurate 
  predictions\cite{xnwang_91} 
  of high-$p_T$ production from $p+A$ collisions.
These calculations can  be extended to the $A+A$ collisions at RHIC. 
In addition, changes in high-$p_T$ production 
  due to the passage of partons through the
  dense environment and a QGP have been predicted\cite{signatures}
  and incorporated into calculations\cite{xnwang_92,kkg}. 
Studying hard probes with the STAR detector will allow measurements
  of how partons are affected in the dense environment and 
  comparisons to pQCD to be made.

\section{The STAR Experiment at RHIC}

During the first year, RHIC will run mainly
 $Au+Au$ collisions and is expected to 
  reach 10\% of the  design luminosity by the end of the year.
The full design luminosity,
  ${\cal L} = 2\cdot 10^{26} ~\rm{cm^{-2}s^{-1}}$, 
  is expected to be attained by the end of year 2.
RHIC will also provide different beam energies and species and will
  include $pA$, $pp$ and polarized $pp$ collisions.


At the heart of the STAR detector\cite{star_detector}
 is the time projection chamber (TPC) 
  enclosed in a $0.5$ Tesla solenoidal magnet.
The TPC covers a range of $|\eta|<2$ over the full azimuth and
 provides charged particle 
 tracking and individual track particle identification
  (PID) for  $p \approx  0.15  - 1.2 ~ \rm{GeV/c}$ and momentum resolution
   $\sigma_p/p \leq 1\%$ for $p< 5 ~\rm{GeV/c}$.
Inside the TPC, is a silicon vertex detector (SVT) 
  which provides tracking and PID near the vertex point. 
The forward TPC's (FTPC) will extend  the tracking coverage to 
  $2.4<|\eta|<4.0$.
A RICH detector\cite{rich_kunde} will
  be installed at STAR for the first three years.
This detector will provide limited angular coverage of 
  $0<\eta<0.3$, $\Delta \phi \approx 20^{\circ}$  and 
 extends PID to $ p \approx 3-5~\rm{GeV/c}$.

Surrounding the TPC is a finely-segmented
 ($0.05 \times 0.05$ in $\Delta \eta \times 
  \Delta \phi$) 
  Pb-scintillator sampling electromagnetic 
  calorimeter (EMC).
A shower-maximum detector is located at $5X_{\circ}$.
The barrel calorimeter 
  covers a range of $|\eta|<1$ with $\Delta \phi =2\pi$.
In year 1, $10\%$ of the EMC will be installed, with $30\%$ added
  each additional year.
An endcap calorimeter, currently under review,
 will cover
 the range $1.05 <\eta<2$ with $\Delta \phi =2\pi$.



The charged particle multiplicity based hardware trigger (L0) covers the range $|\eta|<2$.
A  software trigger (L3), used for enhancing the desired event samples,
  is expected to be ready for year 2 running.

\section{High-$p_T$ Probes of High Energy-Density Matter}


STAR will search 
  for changes in production and correlations 
  of quantities at high-$p_T$ using heavy 
  ion collisions.
Measurements will be done as a function of the
  amount of dense matter traversed  
  by varying the centrality of collisions 
  and using different beam species and energies.
Data from $p+p$ and $p+A$ running will be used 
 as a baseline for the high-$p_T$ measurements.
Some of the proposed signatures of QGP formation 
  that STAR plans to measure are described below.

\begin{figure}[tbh]
  \begin{center}
  \begin{tabular}{lr}
   \epsfxsize=10pc  
       \epsfbox{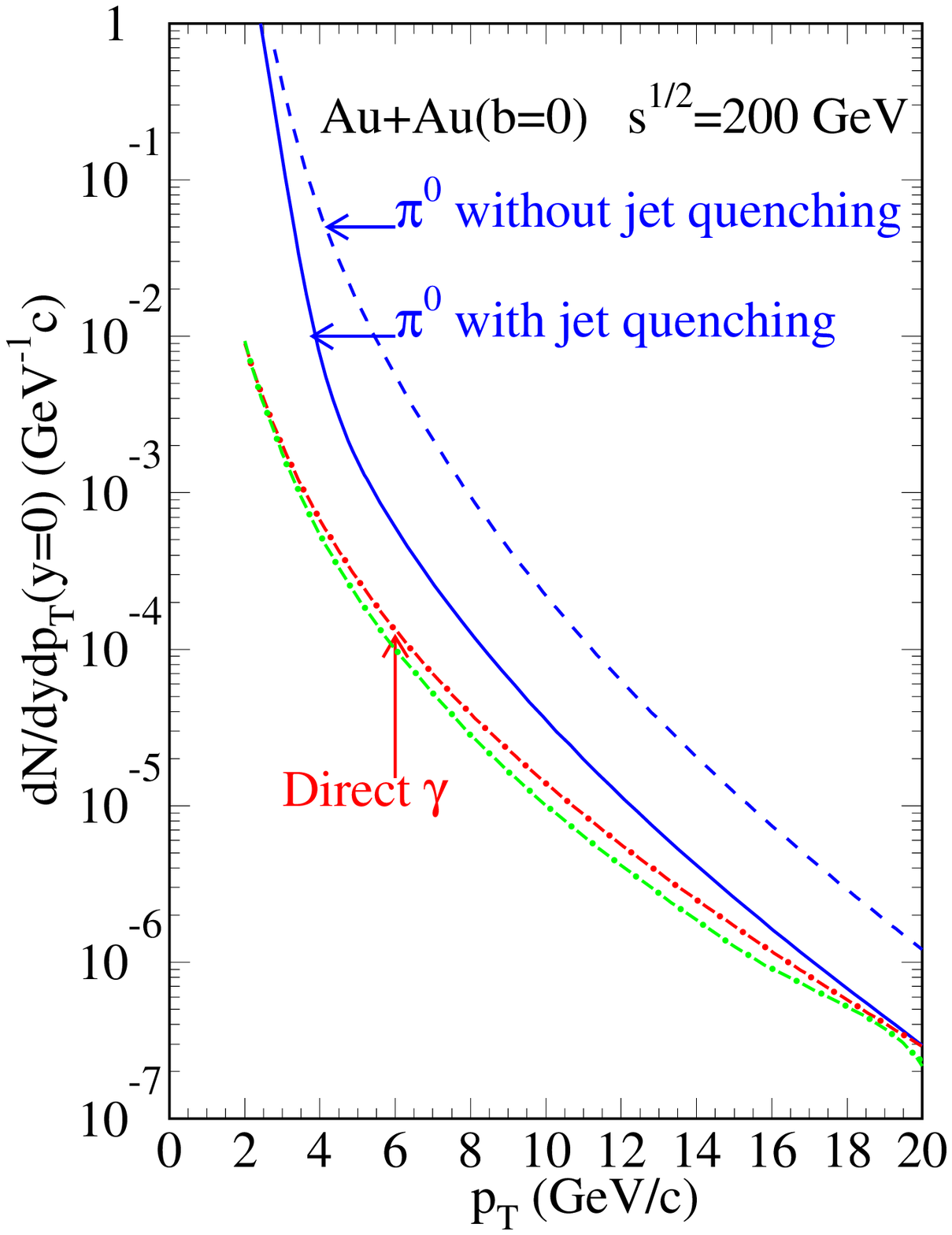} &
   \epsfxsize=18pc  
   \epsfbox{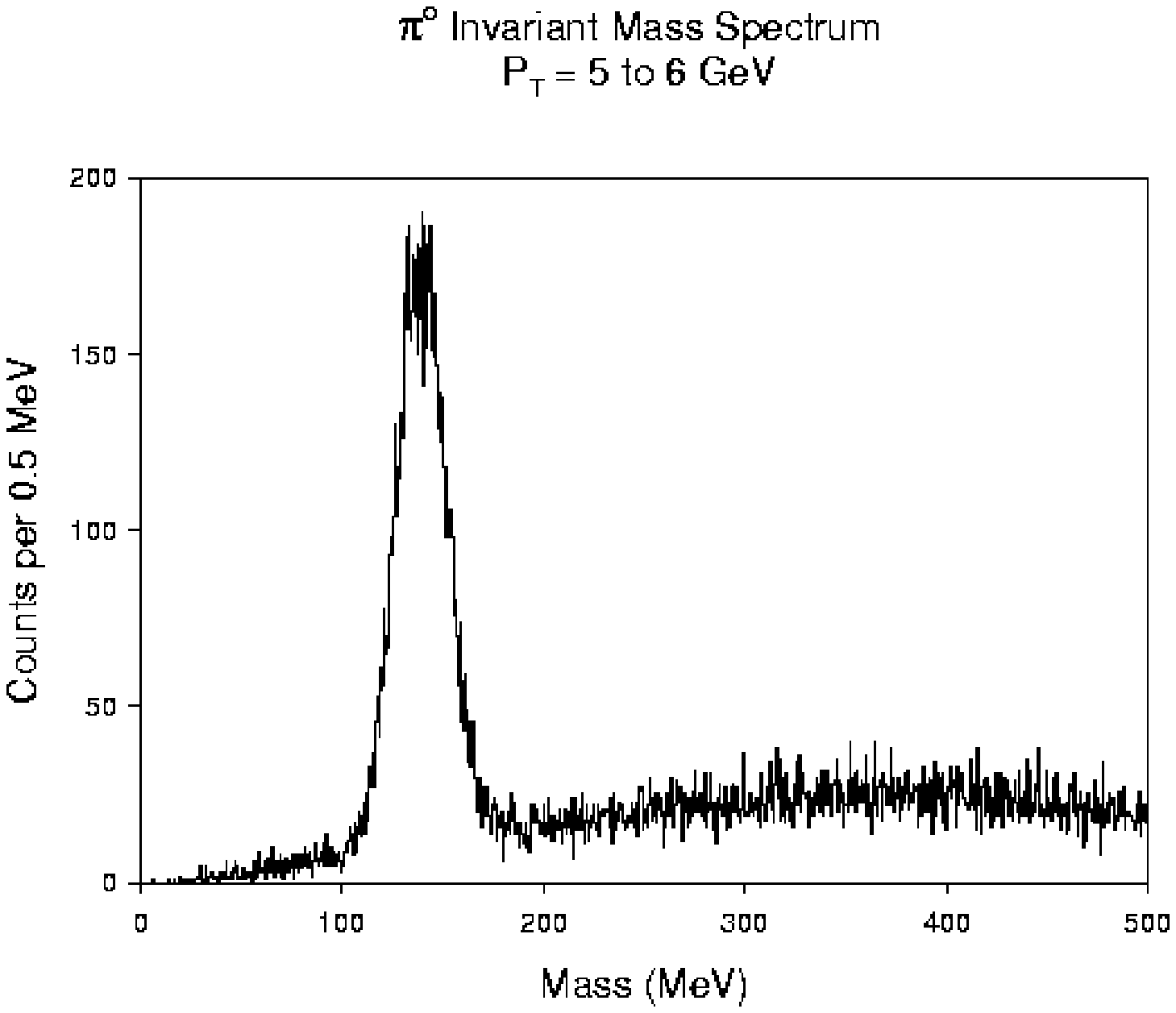}
  \end{tabular}
    \caption{(a) pQCD prediction$^{12}$
     of inclusive $p_T$ spectrum of $\pi^0$'s 
      with and without $dE/dx$ energy loss in $Au+Au$ collisions
      compared to direct $\gamma$'s.  
          (b) Simulation of reconstructed $\pi^0\rightarrow \gamma\gamma$ 
              mass spectrum with year 1 
              statistics in range $p_T = 5$ to $6 ~{\rm GeV/c}$.}
    \label{fig:xnwang_pizero}
  \end{center}
 \end{figure}

A predicted signal of QGP formation using hard probes is 
  ``jet quenching''\cite{jet_quenching},
  which is the softening of the $p_T$ spectrum due to
  partons losing energy, in a $dE/dx$ fashion, when propagating 
  through dense matter. 
The $p_T$ spectrum of $\pi^0$'s is 
  predicted\cite{xnwang_98}
  to soften if there are jet quenching effects in addition to 
  standard nuclear effects,
  as shown by the pQCD prediction in Fig.\ref{fig:xnwang_pizero}a.
STAR can measure this spectrum by identifying 
  $\pi^0 \rightarrow \gamma\gamma$ events using the EMC.
An event sample of $\approx 2500(25)$ events at $p_T = 5(10) ~{\rm GeV/c}$
  is expected in year 1.
A simulation of the reconstructed $\pi^0 \rightarrow \gamma\gamma$ 
  mass peak with year 1 statistics is shown in Fig.\ref{fig:xnwang_pizero}b.
STAR will also measure charged particle high-$p_T$ spectra in the first year.

\vspace{-1.4in}

\begin{figure}[h]
  \begin{center}
  \begin{tabular}{lr}
   \epsfxsize=8pc 
       \epsfbox{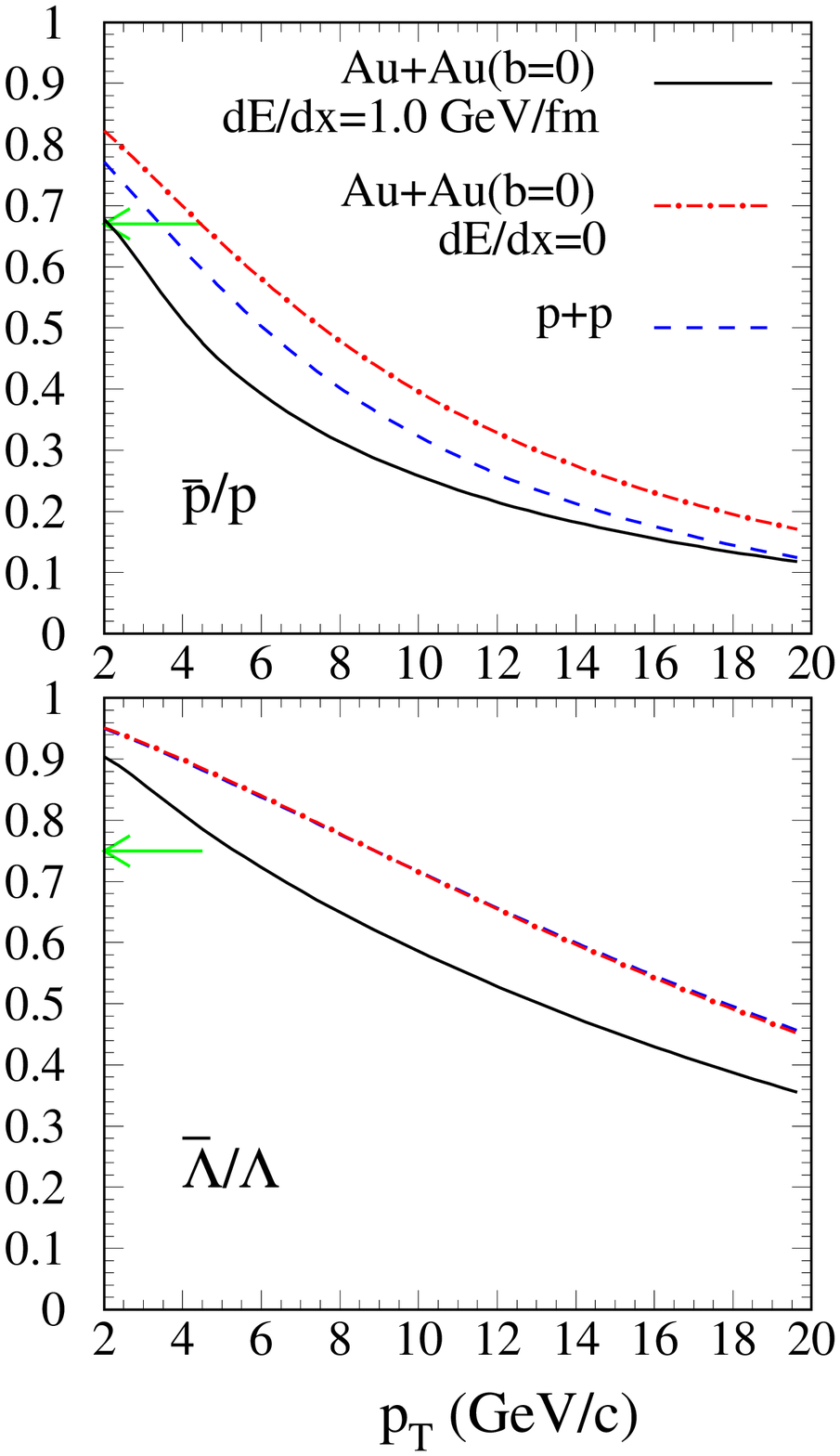} &
   \epsfxsize=12pc 
       \epsfbox{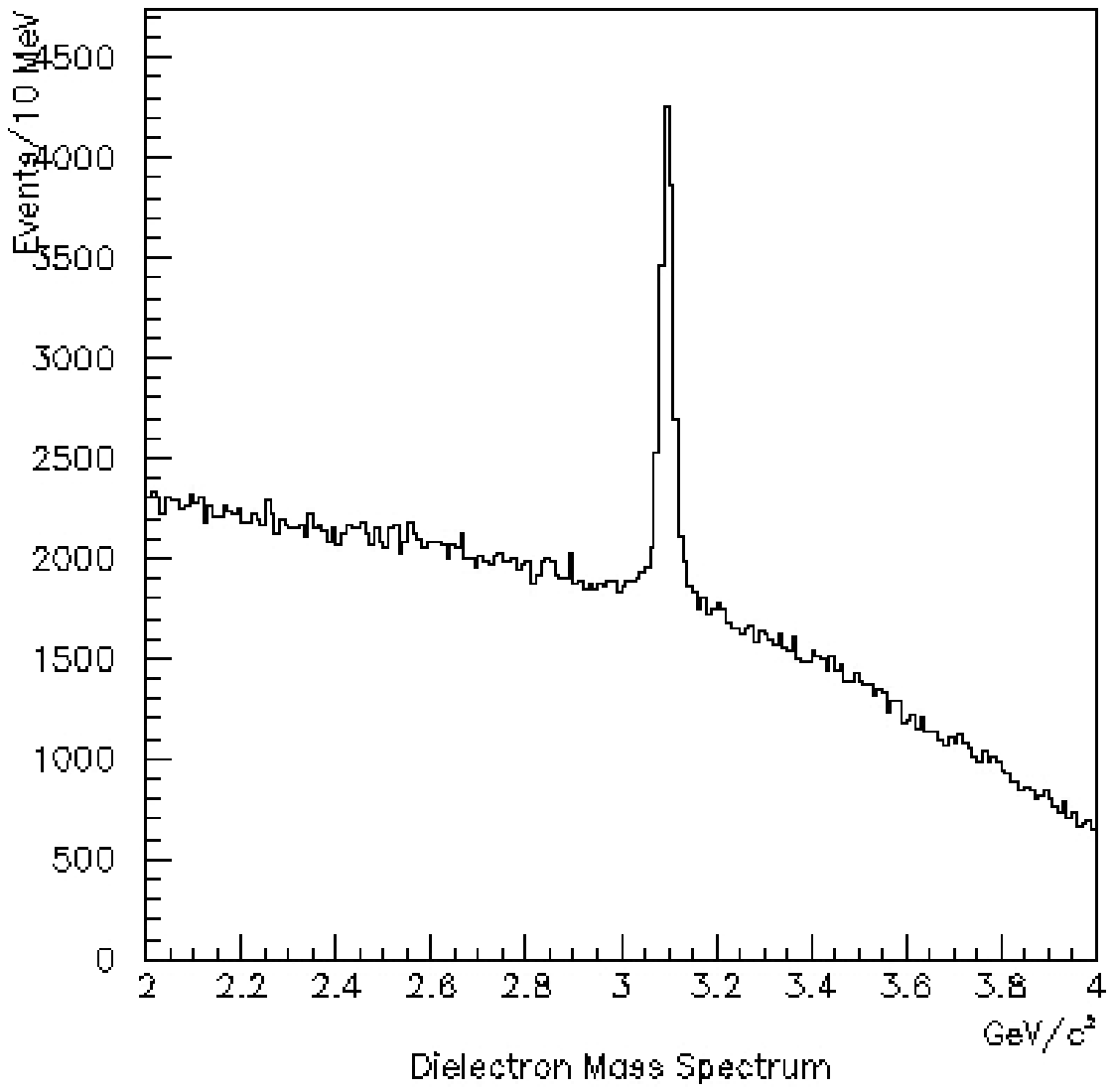} 
   \end{tabular}
    \caption{(a) The predicted$^{12}$ ratio of $\bar p$ to $p$ and $\bar \Lambda$
          to $\Lambda$ spectra as functions of $p_T$ in $pp$
          and central $Au+Au$ collisions with and without 
          energy loss. 
            (b)  The reconstructed $J/\psi\rightarrow e^+e^-$ mass 
               from a full detector simulation, assuming one year of full
               luminosity.}
    \label{fig:ratio_jpsi}
  \end{center}
 \end{figure}

The ratio of charged hadron to anti-hadron
  production as a function of $p_T$ has been predicted\cite{xnwang_98}
  to change for $Au+Au$ relative to $p+p$ 
  collisions as shown in
  Fig.\ref{fig:ratio_jpsi}a.
This ratio changes as a function of the energy loss factor due to jet quenching
  included in the calculation.
The particle dependence of these ratios are due to differences
  in gluon and quark jet quenching in the dense medium.
Using the RICH detector, it will be possible to measure the
  $\bar p/p$ ratio out to $\approx 5 ~ {\rm GeV/c}$ in year 1.


$J/\psi$ production  in a QGP is predicted\cite{jpsi_supp} to be 
  suppressed due to Debye screening of color charges in the
  plasma.
  STAR can measure $J/\psi \rightarrow e^+e^-$ events with
  the inclusion of the L3 trigger and using the TPC 
  and EMC detectors.
The reconstructed $J/\psi$ mass 
  from a simulation is  shown in Fig.\ref{fig:ratio_jpsi}b.
With the full STAR detector and statistics
  expected from one year of running with full  design
  luminosity and a   requirement on the electrons of 
   $p_T > 1.5~\rm{ GeV/c}$,
  STAR can expect to collect  $\approx 4\times 10^4$ $J/\psi$'s 
  per year.

Other high-$p_T$ probes STAR can use are  direct $\gamma$'s and jets.
Due to the large underlying event energies at 
  RHIC, identification
  and energy measurements of jets may be difficult.
Other ways to measure jets may be to use
 leading particle distributions or   $\gamma$+jet events where the $\gamma$ is tagged and used 
  to identify and assign an energy to the jet.
Studies of jets in the heavy ion collisions
  may also include angular correlations and 
  di-jet production.


\section{Summary}

STAR's capabilities for  using hard probes 
  will include  measurements of
  charged, neutral and 
  leading hadron spectra, particle ratios,
 angular correlations, direct $\gamma$'s,  $J/\psi$ and jet production.
Simulations described here
  show the expected results for charged hadron ratios 
  and $\pi^0$ and $J/\psi$ spectra in the first few years of running.
Using data from various energies, 
  beam species and centralities, 
  STAR will be able to provide detailed 
  measurements of high-$p_T$
  production in the dense environment.
The RHIC collider will offer unique and new regimes of dense matter
  and an excellent environment for new physics in the near future. 


\section*{Acknowledgments}
This work was supported in part by U.S. Department of Energy Contract No. DE-AC02-98CH10886.
Special thanks to 
 Tom Cormier, 
 Peter Jacobs, 
 Gerd Kunde, 
 Brian Lasiuk,
 Craig Ogilvie, 
 Jack Sandweiss
 and  Thomas Ullrich
  for help with and
  contributions to this talk.

\end{document}